\documentclass{article}
\usepackage{graphicx}
\usepackage{epsfig}
\usepackage{amsfonts}
\usepackage{enumitem} 
\usepackage{cite}
 \usepackage{amssymb}

\date{\today}

\newcommand{\insertplot}[5]{\begin{figure}
 \hfill\hbox to 0.05in{\vbox to #5in{\vfill
 \inputplot{#1}{#4}{#5}}\hfill}
 \hfill\vspace{-.1in}
 \caption{#2}\label{#3}
 \end{figure}}
 \newcommand{\inputplot}[3]{
 \special{ps: plotfile #1}
}
\newcounter{fig}

\newcommand{\ee}{\end{equation}}
\newcommand{\eea}{\end{eqnarray}}
\newcommand{\be}{\begin{equation}}
\newcommand{\bea}{\begin{eqnarray}}

\begin{document}

\title{{\bf  
Probing the universality of synchronised hair \\ around rotating black holes with $Q$-clouds
}}

\author{
{\large Carlos Herdeiro}$^{1}$,
{\large Jutta Kunz}$^{2}$,
{\large Eugen Radu}$^{1}$
 and
{\large Bintoro Subagyo}$^{3}$
\vspace{0.5cm}
\\
$^{1}${\small Departamento de F\'\i sica da Universidade de Aveiro and CIDMA, Campus de Santiago, 3810-183 Aveiro, Portugal}
   \\
$^{2}${\small Institut f\"ur  Physik, Universit\"at Oldenburg, Postfach 2503 
  D-26111 Oldenburg, Germany}
   \\
$^{3}${\small Department of Physics, Institut Teknologi Sepuluh Nopember, Indonesia
 }
} 
 
 \date{December 2017}
  \maketitle 

\begin{abstract} 
Recently, various families of black holes (BHs) with synchronised hair have been constructed. These are  rotating BHs surrounded, as fully non-linear solutions of the appropriate Einstein-matter model, by  a non-trivial bosonic field in synchronised rotation with the BH horizon. Some families bifurcate \textit{globally} from a bald BH ($e.g.$ the Kerr BH),  whereas others bifurcate only \textit{locally} from a bald BH ($e.g.$ the $D=5$ Myers-Perry BH). It would be desirable to understand how generically synchronisation allows hairy BHs 
to bifurcate from bald ones. 
However, the construction and scanning of the domain of existence of the former families of BHs can be a difficult and time consuming (numerical) task. Here, we first provide a simple perturbative argument to understand the generality of the synchronisation condition. Then, we observe that the study of \textit{Q-clouds} is a generic tool to establish the existence of BHs with synchronised hair bifurcating (globally or locally) from a given bald BH \textit{without} having to solve the fully non-linear coupled system of Einstein-matter equations. As examples, we apply this tool to establish the existence of synchronised hair around $D=6$  Myers-Perry BHs, $D=5$ black rings and $D=4$ Kerr-$AdS$ BHs, where $D$ is the spacetime dimension. The black rings case provides an example of BHs with synchronised hair beyond spherical horizon topology, further establishing the generality of the mechanism. 
\end{abstract}

\section{Introduction and motivation}

Test field analysis on curved spacetimes, and in particular BH geometries, is a useful tool to infer conclusions about more difficult fully non-linear problems. It is instructive to recall a few well known examples.

The Maxwell equations, $\nabla_\mu F^{\mu\nu}=0$, on the Schwarzschild black hole (BH), admit a static, spherical, finite energy, regular solution on and outside the event horizon (see, $e.g.$~\cite{Herdeiro:2015waa}). This solution describes the electric field of a point charge on this geometry and it establishes that -- at least in the neighbourhood of Schwarzschild -- that there are fully-non linear spherical BH solutions of the Einstein-Maxwell theory with electric charge.  This is the well known Reissner-Nordstr\"om BH. By contrast, higher multipoles of the electric field do not have a regular solution \textit{both} on the event horizon \textit{and} at infinity. This hints at the absence of static, asymptotically flat, regular on and outside the event horizon, Einstein-Maxwell BHs with higher electric multipoles (for a single horizon), a fact mathematically established by Israel's uniqueness theorem for electrovacuum~\cite{Israel:1967za}.

In the case of the massive Klein-Gordon equation, $\nabla^2 \Phi=\mu^2\Phi$, on the Schwarzschild BH, there are no regular (on and outside the horizon) static solutions (see, $e.g.$~\cite{Herdeiro:2015waa}). Again, this hints at the absence of time independent scalar hair around a spherical BH. Indeed, an elegant theorem by Bekenstein~\cite{Bekenstein:1972ny} shows there are no stationary, axisymmetric BHs with (stationary, axisymmetric) scalar hair in the Einstein-(massive)-Klein-Gordon theory. Remarkably, however, if the scalar field has a  harmonic dependence in the Killing directions associated to stationarity ($\partial_t$) and axi-symmetry ($\partial_\varphi$), $\Phi\sim e^{-iwt+im\varphi}$ (where $m\in \mathbb{Z}^*$, $w\in \mathbb{R}$), there are finite energy, regular solutions, on and outside the event horizon of the Kerr BH. These are called (linear)  \textit{stationary scalar clouds}~\cite{Hod:2012px} (see also~\cite{Hod:2013zza,Herdeiro:2014goa,Hod:2016lgi}) and establish that -- at least in the neighbourhood of Kerr -- there are fully-non linear BH solutions of the Einstein-(massive and complex)-Klein-Gordon theory.\footnote{The complexity of the scalar field is necessary for the energy-momentum tensor to be compatible with the metric stationarity and axi-symmetry. At the test field level, stationary clouds exist for a real scalar field; but when back reacting, only in the case of a complex scalar field, stationary hairy BHs are possible.} This family of solutions, \textit{Kerr BHs with synchronised scalar hair}, was constructed in~\cite{Herdeiro:2014goa} (see also~\cite{Herdeiro:2015gia,Chodosh:2015oma}) and was generalized to self-interacting scalar fields in~\cite{Kleihaus:2015iea,Herdeiro:2015tia}.

The anticipation of Kerr BHs with scalar hair by using a \textit{linear} test field analysis can be associated to the fact that these BHs bifurcate \textit{globally} from the Kerr solution. This means that there are hairy solutions arbitrarily close to Kerr BHs, both in terms of the local geometry and in terms of the global quantities (ADM mass $M$ and total angular momentum $J$). The hairy BHs meet the Kerr BHs 2-parameter domain of existence in a 1-dimensional subset, called~\textit{an existence line}, whose precise location within the Kerr domain of existence depends on the  winding number $m$ of the hairy BH solutions. An analogous global bifurcation is found for other BHs with synchronised hair, namely Kerr BHs with Proca hair~\cite{Herdeiro:2016tmi,Herdeiro:2017phl}, Kerr-Newman BHs with scalar hair~\cite{Delgado:2016jxq} and $D=5$ Myers-Perry (MP)-$AdS$ BHs with scalar hair~\cite{Dias:2011at}. Global bifurcation of a family of BHs with synchronised hair from a family of bald BHs is intimately related to the existence of a superradiant instability for the bald BHs, triggered by the field that endows the BHs with hair~\cite{Herdeiro:2014ima}.

There are also, however, families of BHs with synchronised hair that do not bifurcate globally from a family of bald BHs, but only \textit{locally}. This is the case of the asymptotically flat MP BHs with synchronised hair found in~\cite{Brihaye:2014nba,Herdeiro:2015kha}, in $D=5$. This means that there are hairy solutions with a horizon geometry that is arbitrarily close to that of a MP BH, but the global quantities are not. There is always a \textit{mass gap}: even a very dilute scalar field outside the horizon, that has an arbitrarily small local deformation of the corresponding MP geometry, integrates to give a finite energy that never approaches zero. In this case, the existence of synchronised hair around the MP BH could not be anticipated by solving the linear, massive, Klein-Gordon equation on this background, since it has no linear stationary  cloud solutions. Equivalently there is (at least in general) no existence line for zero modes of the superradiant instability, because there is no superradiant instability of the asymptotically flat MP BH triggered by a linear Klein-Gordon field. 

\bigskip 

One may ask nonetheless, if there is any test field analysis that could anticipate the existence of MP BHs with synchronised scalar hair. In this paper we claim there is, and that the corresponding probes, hereafter dubbed \textit{Q-clouds}, following~\cite{Herdeiro:2014pka}, are a useful tool in establishing the existence of synchronised hair, for generic bald BHs. 

$Q$-clouds are the counterpart on a BH geometry of the well known Minkowski spacetime \textit{Q-balls}~
\cite{Friedberg:1976me,Coleman:1985ki}. The latter are solitonic solutions on Minkowski spacetime, that exist for non-gravitating complex scalar fields with some types of self-interactions. Both spherical and rotating 
solutions exist, assuming again a scalar field with form $\Phi\sim e^{-iwt+im\varphi}$, and they are confined to a frequency domain, $w_{\rm min}<w<w_{\rm max}$. At both ends of this frequency domain the solutions diverge in their physical quantities (mass and angular momentum). In~\cite{Herdeiro:2014pka} it was observed that rotating $Q$-balls survive when replacing the Minkowski background by a Kerr BH geometry and imposing the synchronization condition, becoming $Q$-clouds. These solutions exist on a 2-dimensional open subset of the 2-dimensional parameter space of Kerr BHs. Again, they are confined to a frequency range $w_{\rm min}<w<w_{\rm max}(M)$.  $w_{\rm min}$ is the same as for rotating $Q$-balls (with the same potential), and $Q$-clouds still appear to diverge in this limit, on the BH background; but their behaviour at  $w_{\rm max}(M)$ is quite different. As $w\rightarrow w_{\rm max}(M)$, which now depends on the BH mass $M$, $Q$-clouds decrease their amplitude and become the stationary clouds of linear Klein-Gordon theory (the self interactions become negligible). The curve $w_{\rm max}(M)$ becomes the aforementioned existence line (for the appropriate value of $m$).

$Q$-clouds are therefore scalar bound states around Kerr BHs which, generically, 
are not zero modes of the superradiant instability: they \textit{do not rely} on the existence of these zero modes. But $Q$-clouds reduce to these zero modes when approaching the corresponding existence line: they \textit{are sensitive} to these zero modes. This dichotomy makes $Q$-clouds a generic probe of synchronised scalar hair. If a BH background admits regular synchronised $Q$-cloud solutions, it can support scalar hair -- at least close to the bald solution -- and if the $Q$-clouds reduce to linear clouds at some boundary of their domain of existence one even identifies a global bifurcation of the bald BH towards a family of BHs with synchronised hair. Using this rationale, 
in this paper we generalise the results in~\cite{Herdeiro:2014pka} by replacing the Kerr BH background by its natural higher dimensional generalisation: a MP BH rotating in a single plane. We show $Q$-clouds exist, confirming the results in~\cite{Brihaye:2014nba,Herdeiro:2015kha} for $D=5$ and anticipating MP BHs with synchronised hair also exist in $D\geqslant 6$. We also find $Q$-clouds for a $D=5$ black ring background, showing BHs with synchronised hair exist beyond spherical horizon topology, and also for $D=4$ Kerr-$AdS$, to establish this tool also works in non-asymptotically flat spacetimes. We emphasise that  the study of $Q$-clouds is technically much simpler but it contains already the basic ingredients of the solutions to the full Einstein-Klein-Gordon system. 

This paper is organised as follows. In Section~\ref{sec2} we establish the generic background geometry and scalar field model we shall work with. In Section~\ref{sec3} we present a general argument on the validity of the synchronisation condition, to allow non-trivial scalar field configurations on the geometries we are considering. In Section~\ref{sec4} we exhibit the results for the different BH backgrounds considered in this paper (and also higher dimensional flat space).  In Section~\ref{sec5} we present some final remarks. We use units with $c=1=G$.

\section{The framework}
\label{sec2}

\subsection{The background geometry}
The background geometry we shall consider describes a generic $D$ dimensional BH solution: {\bf (i)} with a single non-degenerate horizon, which needs not be topologically spherical; {\bf (ii)}  which is regular on and outside the horizon; {\bf (iii)} which is asymptotically flat or $AdS$; {\bf (iv)} 
which rotates in a single plane. Thus, we take the following generic line element form, in $D\geqslant 4$ dimensions
 \begin{eqnarray}
\label{metric-g}
 ds^2=-F_0(r,\theta) dt^2+F_1(r,\theta)
\left[
dr^2 +\Delta(r) d\theta^2
\right]
+F_2(r,\theta) d\Omega_{D-4}^2+F_3(r,\theta) [d\varphi-W(r,\theta) dt]^2\ .
\end{eqnarray}
This metric is given in terms of five metric functions, 
$F_i(r,\theta)$ (with $i=0,1,2,3$) and $W(r,\theta)$ 
whose explicit expressions depend on the specific solutions to be considered below.
$\Delta(r)$ is a function to be conveniently chosen, by using the residual metric gauge freedom. $d\Omega_{n}^2$ is the metric on the $n-$dimensional sphere, $\mathcal{S}^n$, and $r, t$ are the radial and time coordinates, respectively.
The range of the radial coordinate is $r_H\leqslant r<\infty$;  
$\varphi$ and $\theta$ are angular variables, with $0\leqslant \varphi <2\pi$,
and $0\leqslant \theta \leqslant \pi/2$; $r=r_H$ corresponds to the event horizon,  
wherein $F_0(r_H)=0$.  

Two particularly relevant Killing vectors are 
 $\xi=\partial_t$ and $\eta=\partial_\varphi$, since $r=r_H$
is a Killing horizon of a combination thereof: 
$\chi=\xi+\Omega_H \eta$, where $\Omega_H$ is the event horizon velocity, $\Omega_H=-{\xi^2}/{\xi \cdot \eta}=-({g_{tt}}/{g_{t\varphi}})|_{r_H}$.
 The mass $M$ and angular momentum  $J$ 
of the background BH are read off from the large $r$ 
 asymptotics of the metric functions,
 $g_{tt}=-1+{C_t}/{r^{D-3}}+\dots,
~g_{\varphi t}=-F_3W= { C_\varphi}\sin^2 \theta/{r^{D-3}}+\dots,$
 with $16 \pi M={(D-2)V_{(D-2)}}C_t$,  $ 8\pi J={V_{(D-2)}}C_{\varphi}$,
where $V_{(p)}$ is the area of $\mathcal{S}^p$. 

We shall present results for two different classes of vacuum
BH solutions and one class of asymptotically $AdS$ solutions, all of which can be described
by~(\ref{metric-g}).
These are: 
 ${\bf i)}$ MP BHs~\cite{Myers:1986un}, which possess a $\mathcal{S}^{D-2}$
horizon topology;  
${\bf ii)}$
black rings~\cite{Emparan:2001wn,Emparan:2006mm,Kleihaus:2012xh,Kleihaus:2014pha}, with a $\mathcal{S}^1\times \mathcal{S}^{D-3}$ horizon topology;  ${\bf iii)}$ the Kerr-$AdS$ BH.
Non-vacuum generalizations of these solutions can  also be described within this framework, 
in particular the configurations with scalar hair that are suggested by the existence of the $Q$-clouds considered herein.

Following the usual convention in the (higher dimensional) BH
literature~\cite{Emparan:2008eg}, we fix the overall scale 
of the solutions by fixing their mass $M$.
Then one defines a reduced angular momentum
%
%
\begin{eqnarray}
\label{dim1}
j\equiv c_j \frac{J}{{  M}^{\frac{D-2}{D-3}}} \ , 
~~~{\rm with}~~~~ c_j\equiv  \frac{1}{\sqrt{D-3}}
\left(
\frac{\pi^{(D-2)/2} (D-2)^{D-2}}{2^{D}\Gamma(\frac{D-2}{2})}
\right)^{\frac{1}{D-3}}~.
 \end{eqnarray}
 %

%
\subsection{The scalar field }

The scalar test field that we shall consider on the BH backgrounds described in the previous subsection is complex and with a self-interaction potential, in $D$ spacetime dimensions. Its dynamics follows from the action
\begin{equation}
\label{action}
S=-\int \left[ 
   \frac{1}{2} g^{\mu\nu}\left( \Phi_{, \, \mu}^* \Phi_{, \, \nu} + \Phi _
{, \, \nu}^* \Phi _{, \, \mu} \right) + U( \left| \Phi \right|) 
 \right] \sqrt{-g} d^Dx
\ , 
\end{equation}
where  
the asterisk denotes complex conjugation, $\Phi_{,\, {\mu}}  \equiv {\partial \Phi}/{ \partial x^{\mu}}$,
and $U$ is the scalar potential.
Variation with respect to the scalar field
leads to the Klein-Gordon equation 
\begin{equation}
\label{KG}
\left(\nabla^2-\frac{\partial U}{\partial\left|\Phi\right|^2}\right)\Phi =0 ~.
\end{equation} 
Our choice for the scalar potential is a standard one in the $D=4$ $Q$-balls literature:
\begin{equation} 
U(|\Phi|) =  \mu^2 |\Phi|^2-\lambda |\Phi|^4 +\beta |\Phi|^6 \ ,
\label{U} 
\end{equation}
where $\mu$ is the  boson mass. This potential allows the existence of non-topological soliton-type solutions
 on Minkowski spacetime ($Q$-balls). It has a minimum, $U(0)=0$, at $\Phi =0$
and a second minimum at some finite value of $|\Phi|$.
In practice, following the $D=4$~\cite{Volkov:2002aj,Kleihaus:2005me,Radu:2008pp}
 literature, the numerical results reported in this work 
use the following potential parameters 
\begin{equation}
\beta = 1,~~\lambda=2,~~\mu^2=1.1 \ .
\label{param}
\end{equation}
Also, all scalar field quantities are given in units set by the mass parameter $\mu$.
 
Variation of (\ref{action}) with respect to the metric tensor $g_{\mu\nu}$
leads to the scalar field energy-momentum tensor
\begin{eqnarray}
T_{\mu \nu} =  \Phi_{, \, \mu}^*\Phi_{, \, \nu}
+\Phi_{, \, \nu}^*\Phi_{, \, \mu} 
-g_{\mu\nu} \left[ \frac{1}{2} g^{\alpha\beta} 
\left( \Phi_{, \, \alpha}^*\Phi_{, \, \beta}+
\Phi_{, \, \beta}^*\Phi_{, \, \alpha} \right)+U( \left| \Phi \right|)\right]~.
\label{tmunu} 
\end{eqnarray}

A scalar field ansatz which yields an energy-momentum tensor compatible with
the symmetries of the line element (\ref{metric-g}) reads
\begin{eqnarray}
\label{scalar-ansatz}
\Phi=\phi(r,\theta)e^{i(m \varphi-w t)}
\end{eqnarray}
 where $\phi$ is a real function defining the scalar field profile, $w>0$ is the frequency and $m\in \mathbb{Z}$
is an azimuthal winding number.

With the ansatz~(\ref{scalar-ansatz}), the Klein-Gordon eq. (\ref{KG}) for the metric~(\ref{metric-g}) is, explicitly:
\begin{eqnarray}
\label{eqZ}
\phi''+\frac{1}{\Delta}{\ddot \phi}
+\frac{1}{2}
\left[
\frac{\Delta'}{\Delta}
+\frac{F_0'}{F_0}
+\frac{(D-4)F_2'}{F_2}
+\frac{F_3'}{F_3}
\right]
\phi'
+
\frac{1}{2\Delta}
\left[
\frac{\dot F_0}{F_0}
+\frac{(D-4)\dot F_2}{F_2}
+\frac{\dot F_3}{F_3}
\right]
\dot \phi
\\
\nonumber
+\frac{F_1(w-m W)^2}{F_0}\phi
-\frac{m^2F_1}{F_3}\phi
-F_1( \mu^2-2\lambda \phi^2+3 \beta \phi^4)\phi=0~,
\end{eqnarray}
where a prime denotes $\partial_r$, and a 
dot denotes $\partial_\theta$.
The energy and angular momentum densities of the field 
are
\begin{eqnarray}
-T_t^t=
\frac{1}{F_1}\left(\phi'^2+\frac{\dot \phi^2}{\Delta}\right)
+
\left(
\frac{m^2}{F_3}
+\frac{w^2-m^2 W^2}{F_0}
+\mu^2-\lambda\phi^2+\beta \phi^4
\right)\phi^2\ ,
\qquad
T_\varphi^t=\frac{2m (w-m W)}{F_0}\phi^2.
\end{eqnarray}
 
The total energy $E_{(\Phi)}$ and angular momentum $J_{(\Phi)}$ of the $Q$-clouds to be discussed below are then
\begin{eqnarray}
\label{scalar-charges}
E_{(\Phi)}=- 2\pi V_{D-4} \int_{r_H}^\infty dr \int_0^{\pi/2} d\theta \sqrt{-g}T_t^t \ , \qquad 
J_{(\Phi)}= 2\pi V_{D-4}  \int_{r_H}^\infty dr \int_0^{\pi/2} d\theta \sqrt{-g}T_\varphi^t~,
 \end{eqnarray}
with
$\sqrt{-g}=F_1\sqrt{F_0F_3\Delta}F_2^{(D-4)/2}$. 

A 
 conserved \textit{Noether} charge $Q$ is associated with the complex scalar field $\Phi$, since the Lagrange density is invariant under
the global phase transformation $\Phi \to \Phi e^{i\alpha}$, leading to the conserved current $j^{\mu}=-i\left[
\Phi^* \partial^{\mu}\Phi-\Phi \partial^{\mu}\Phi^*
\right]$, with $\nabla_\mu j^{\mu}=0$,
and a conserved charge $Q$, given by the integral of $j^t$ on a space-like surface.
One can easily see, that in the absence of backreaction,
the following relation holds: $J_{(\Phi)}=m Q$.
Thus,  angular momentum  is quantised for any value of $D$.

We remark that the $D=4$ $Q$-clouds on a Kerr background~\cite{Herdeiro:2014pka}
 can also be studied within this framework,
in which case one takes 
$F_2=1$,  $0\leqslant \theta \leqslant \pi $ and $V_{D-4}=1$ in the above relations
(also $j=J/M^2$ for $D=4$).

\section{The synchronization condition}
\label{sec3}

We assume that the metric functions in the generic line element (\ref{metric-g}) have a power series expansion as $r\to r_H$. Recalling that we are focusing here on non-degenerate horizons, we have:
 \begin{eqnarray}
\label{rh}
&&
F_0(r,\theta)=f_0^{(2)}(\theta)(r-r_H)^2+f_0^{(3)}(\theta)(r-r_H)^3+\dots,
\\
\nonumber
&&
F_i(r,\theta)=f_i^{(0)}(\theta)+f_i^{(1)}(\theta)(r-r_H) +f_i^{(2)}(\theta)(r-r_H)^2+ \dots, \qquad i=1,2,3 \ ,
\\
\nonumber
&&
W(r,\theta)=\Omega_H+w_2(\theta)(r-r_H)^2+\dots \ .
\end{eqnarray}
For all BHs we shall consider, the residual metric gauge freedom in
 (\ref{metric-g}) is fixed 
by taking
 \begin{eqnarray} 
\label{Delta}
 \Delta(r)=r^2.
 \end{eqnarray} 
The precise expressions of the functions $f_i^{(k)}(\theta)$, $w_k(\theta)$
depend  on the specific solution and it is not important for the argument here. We remark, however, that $f_i^{(0)}(\theta)$ are strictly positive functions.

We also assume that the scalar amplitude $\phi$ possesses a Taylor series expansion
close to the horizon:
 \begin{eqnarray}
\label{rhZ}
\phi(r,\theta)=\phi^{(0)}(\theta)+\phi^{(1)}(\theta)(r-r_H) +\phi^{(2)}(\theta)(r-r_H)^2+ \dots \ . \end{eqnarray}
Then, replacing~(\ref{rhZ}) in the Klein-Gordon equation (\ref{eqZ}), we find, to lowest order, the condition
 \begin{eqnarray}
\label{cond1}
\frac{(w-m\Omega_H)^2f_{1}^{(0)}(\theta)}{f_{0}^{(2)}(\theta)}\frac{\phi(r_H,\theta)}{(r-r_H)^2}=0 \ .
\end{eqnarray}
The first possibility to satisfy this condition is to assume $w-m\Omega_H \neq 0$ and to impose that the scalar field vanishes on the horizon $\phi(r_H,\theta)=0$, $i.e.$, $\phi^{(0)}(\theta)$. 
This implies, however, that the coefficients $\phi^{(i)}$ in (\ref{rhZ})
vanish order by order, and thus  the scalar amplitude $\phi$ is identically zero.
Thus, to have a non-trivial scalar field, we must satisfy condition (\ref{cond1}) by taking:
 \begin{eqnarray}
\label{synch}
\frac{w}{m}=\Omega_H \ ,
\end{eqnarray}
which is {\it the synchronization condition}. We can interpret it as a synchronisation of the angular phase velocity of the scalar field and the angular velocity of the BH horizon. This condition allows a non-vanishing value of the scalar amplitude
at the horizon $\phi(r_H,\theta)\neq 0$ and a non-trivial scalar field profile outside the BH. 

The next order relation reads
 \begin{eqnarray}
\label{synch2}
\frac{\phi^{(1)}(\theta)}{r-r_H}=0,~~i.e.~~~\partial_r \phi\big|_{r=r_H}=0~.
\end{eqnarray}
It follows that the functions $\phi^{(i)}(\theta)$  in (\ref{rhZ}) (with $i\geqslant 2$)
are  determined by 
the scalar field at the horizon $\phi^{(0)}(\theta)$, together with its angular derivatives, 
and the metric functions of the background geometry, $e.g.$:
 \begin{eqnarray}
\label{z2}
\nonumber
\phi^{(2)}=-\frac{1}{4r_H^2}
\left\{
 \ddot \phi^{(0)}
+
\left[
\frac{\dot f_0^{(2)}}{f_0^{(2)}}
+\frac{(D-4)\dot f_2^{(0)}}{f_2^{(0)}}
+\frac{\dot f_3^{(0)}}{f_3^{(0)}}
\right]\frac{\dot \phi^{(0)}}{2}
-
\left[
\frac{m^2}{ f_3^{(0)}}+\mu^2-2\lambda(\phi^{(0)})^2+3\beta (\phi^{(0)})^4
\right]
 f_1^{(0)} \phi^{(0)} r_H^2
\right\}~.
\end{eqnarray}
Observe that neither the horizon topology nor the spacetime dimension
(or far field asymptotics) 
are relevant in  the above discussion.

Finally, one remarks that the condition (\ref{synch})
implies the following (finite) expressions
of the energy density and the angular momentum density
 as $r\to r_H$:
 \begin{equation}
\label{roj}
-T_t^t=
\bigg(
\mu^2
+\frac{m^2}{f_3^{(0)}}-\frac{2m^2 \Omega_H w_2}{f_0^{(2)}}
\bigg)
(\phi^{(0)})^2
-\lambda (\phi^{(0)})^4+\beta (\phi^{(0)})^6
+\frac{(\dot \phi^{(0)})^2}{r_H^2 f_1^{(0)}}+\dots,~~
T_\varphi^t=-\frac{2m^2 w_2 (\phi^{(0)})^2 }{f_0^{(2)}}+\dots~.
\end{equation}
In the case of a degenerate horizon, a similar, albeit more involved analysis, can be performed.   

\section{$Q$-clouds on BH backgrounds}
\label{sec4}

\subsection{Flat spacetime $Q$-balls}
\label{sec_flat}

Before discussing the $Q$-balls on a BH background, it is useful to consider first flat spacetime $Q$-balls.
These are solutions to the model~(\ref{action}) on Minkowski spacetime, describing localised lumps of energy. 
In  $D=4$, $Q$-balls have been extensively discussed in the literature, following the original works~\cite{Friedberg:1976me,Coleman:1985ki}.
They circumvent the standard Derrick-type obstruction for solitons due to having a
time-dependent phase in the scalar field. 
Higher dimensional spherically symmetric $Q$-balls, on the other hand,  are discussed 
$e.g.$ in~\cite{Gleiser:2005iq,Tsumagari:2008bv}.
To the best of our knowledge, however, no study of the properties of spinning $Q$-balls in $D\geqslant 5$ exists.\footnote{
See, however, Ref. \cite{Hartmann:2010pm} for a study of spinning  $D=5$ Q-balls with two equal angular momenta in a model with a complex doublet scalar field.}
Such configurations can easily be studied within the framework of Section 2.
The scalar ansatz is still given by (\ref{scalar-ansatz}) and Minkowski spacetime corresponds to
the following choice in (\ref{metric-g}): $F_0=F_1=1,F_2=r^2\cos^2 \theta, F_3=r^2\sin^2 \theta$ and $\Delta=r^2$.
The Klein-Gordon equation (\ref{KG})
is solved with the following boundary conditions: $
\phi|_{r=0}=\phi|_{\theta=0}=\phi|_{r=\infty}=0,~~\partial_\theta \phi|_{\theta=\pi/2}=0$, 
which follow from regularity requirements.
In our approach, $(w,m)$ and the constants $(\mu,\lambda,\beta)$ in the potential 
are input parameters,
the energy and angular momentum being computed
from the numerical output.

\begin{figure}[t!]
\hbox to\linewidth{\hss%
	\resizebox{8cm}{6cm}{\includegraphics{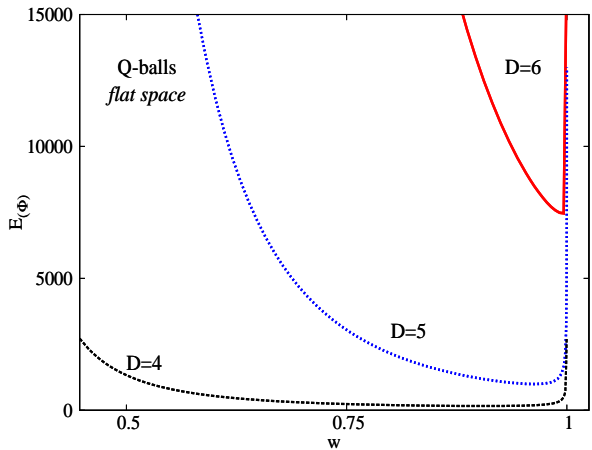}}
\hss}
\caption{\small 
Minkowski $Q$-balls energy $vs.$ scalar field frequency, in $D=4$ (black dashed line), $D=5$ (blue dotted line) and $D=6$ (red solid line).   
}
 \label{flat}
\end{figure}

We have found that the basic features of the  well-known $D=4$ $Q$-balls are shared by their higher dimensional
counterparts.
 For any $D$, the
$Q$-balls 
only exist  within a certain frequency range, 
$w_{\rm min} < w < w_{\rm max}$.
The lower limiting frequency, $w_{\rm min}$, is
determined by the properties of the potential and decreases with $D$; the higher limiting frequency is always 
 $w_{\rm max}=\mu$.
At a critical value of the frequency $w=w_c$, both the energy and the angular momentum of the $Q$-balls reach a minimum, 
whence they monotonically rise towards both limiting values
of the frequency.\footnote{  
This
behaviour remains qualitatively the same for any $m$, in particular for $m = 0$.
}
These behaviours can be observed in Figure~\ref{flat}.  
Rotating $Q$-balls do not possess a slowly rotating limit and are expected to be stable along the lower branch -- from $w_{\rm min}$ to  $w_c$, 
wherein their energy is smaller than that of a collection of free bosons with the same Noether charge.
As $w$ approaches either of the limiting frequencies of the allowed range, the energy and angular momentum grow (apparently) without bounds.
 
 {\small \hspace*{3.cm}{\it  } }

 {\small \hspace*{3.cm}{\it  } }
\begin{figure}[t!]
\hbox to\linewidth{\hss%
	\resizebox{8cm}{6cm}{\includegraphics{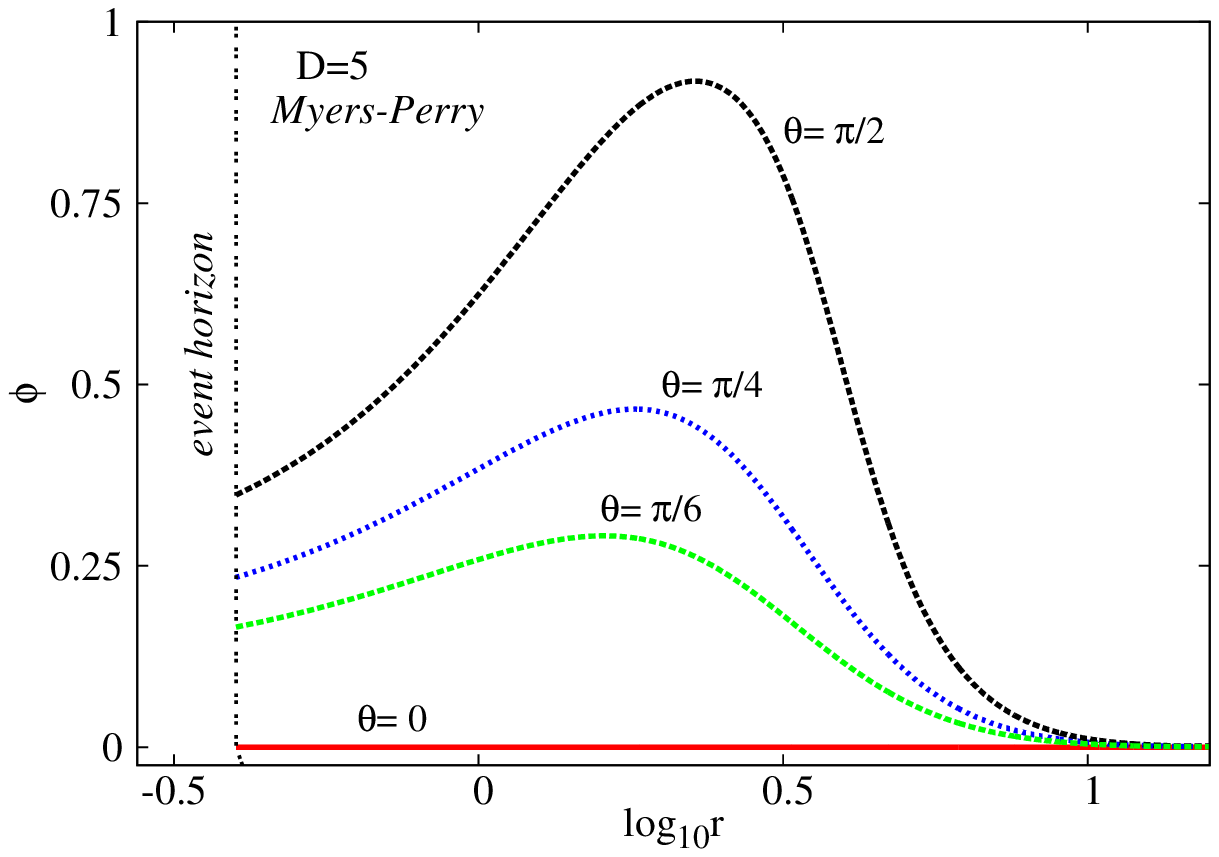}}
	\hspace{10mm}%
		\resizebox{8cm}{6cm}{\includegraphics{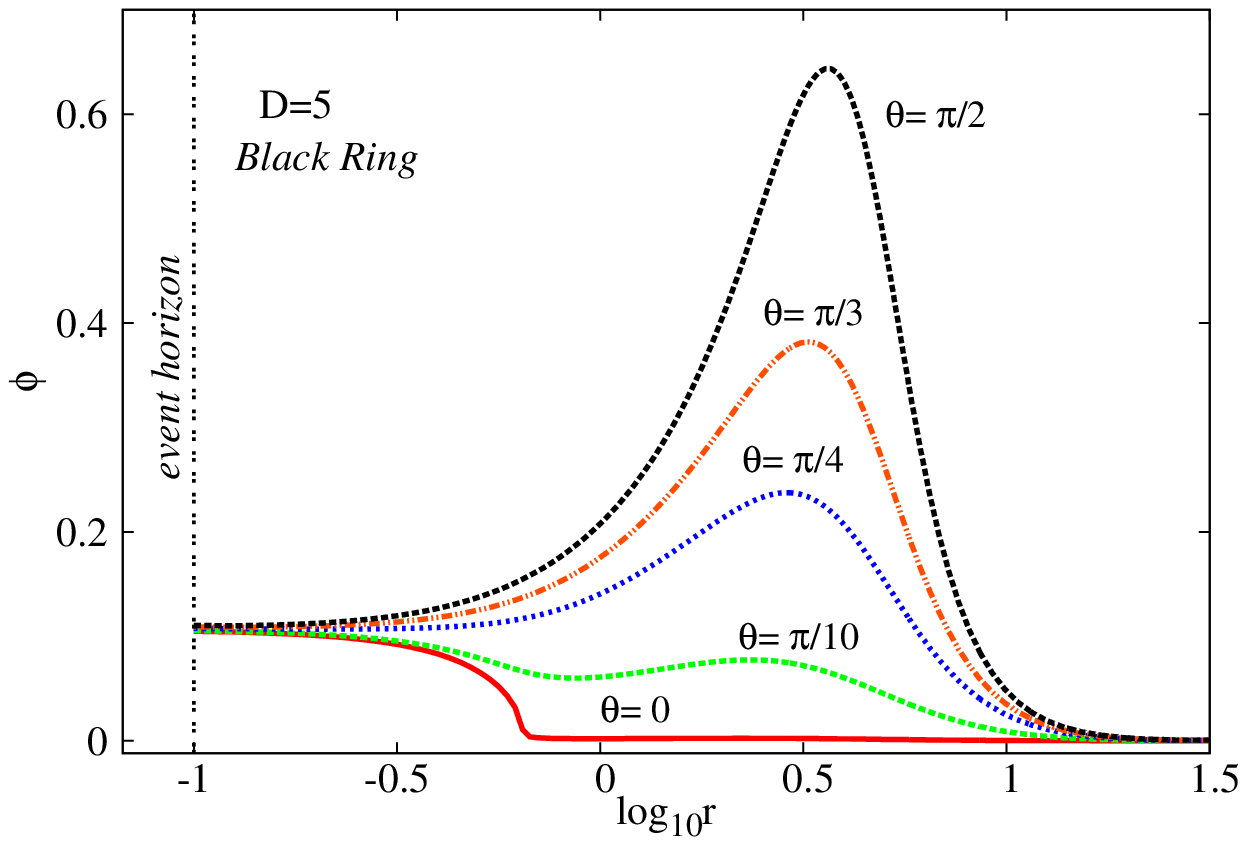}}
\hss}
\caption{\small  
{\it Left:} The radial scalar field profile for several different values
of $\theta$, 
for a typical $Q$-cloud on a $D=5$ MP background with
$r_H=   0.4$ and $a=0.8$. 
{\it Right:} 
Same for a  black ring (BR) background with event horizon radius
$r_H=0.1$ and $S^1$ radius  $R=0.6481$. 
These are the constants entering the BR parametrization in~\cite{Kleihaus:2012xh,Kleihaus:2014pha}; 
$(r,\theta)$ are also the coordinates therein. 
}
\label{profile-MP1}
\end{figure}
 
\subsection{$Q$-clouds on a Myers-Perry BH}

The first example of a BH background is the MP BH in $D=5,6$. The metric functions in (\ref{metric-g}) are\footnote{
When written in this form, the near horizon expansions of the MP solution differ from those in (\ref{rh}).
The relations (\ref{rh}) are recovered in a different radial coordinate.  However, the BH solution is more cumbersome in that case.
} \cite{Emparan:2008eg}:
\begin{eqnarray}
\nonumber
&&
F_0(r,\theta)=\frac{\Delta(r)}{(r^2+a^2)P(r,\theta)} \ , \qquad
F_1(r,\theta)=\frac{r^2+a^2 \cos^2\theta}{\Delta(r)} \ , \qquad
F_2(r,\theta)=r^2\cos^2\theta ,
\\
&&
F_3(r,\theta)= P(r,\theta)(r^2+a^2)\sin^2\theta \ , \qquad
W(r,\theta)=\frac{M}{r^{D-5 }}\frac{a}{(r^2+a^2) (r^2+a^2\cos^2\theta)P(r,\theta)}~,
\end{eqnarray}
where
\begin{eqnarray}
\Delta(r)=(r^2+a^2)
\left[1-\frac{U}{r^{D-5}(r^2+a^2) }
\right] \ , \qquad 
P(r,\theta)=1+\frac{U}{r^{D-5}}\frac{a^2\sin^2\theta}{(r^2+a^2) (r^2+a^2\cos^2\theta) } \ ,
\end{eqnarray}
and $U,a$ are two input parameters. 
These BHs have a horizon of spherical topology located
at 
$r=r_H$, where $\Delta(r_H)=0$, which implies
$
U=(r_H^2+a^2) r_H^{D-5}.
$
%
The quantities of interest for our study are
\begin{eqnarray}
M=\frac{(D-2)V_{(D-2)}r_H^{D -5}}{16\pi }(r_H^2+a^2) \  , \qquad
J=\frac{V_{(D-2)}}{8\pi}a r_H^{D -5}(r_H^2+a^2) \ , \qquad
\Omega_H=\frac{a}{a^2+r_H^2} \ .
\end{eqnarray}

 Thus the dimensionless angular momentum, (\ref{dim1}), is: 
\begin{eqnarray}
\label{MP-rel1}
\nonumber
&&
j  =\frac{1}{\sqrt{D-3}} 
\left(
\frac{\sqrt{\pi}\Gamma(\frac{D-1}{2})}{\Gamma(\frac{D-2}{2})}
\right)^{\frac{1}{D-3}}
\frac{x}{(1+x^2)^{\frac{1}{D-3}}}\ ,~~{\rm with} ~~~x\equiv \frac{a}{r_H},~~0\leqslant x<\infty~.
\end{eqnarray} 
The only case where
extremality is possible (which is reached for $j\to 1$) 
is $D=5$, 
a limit which, however, is singular. 
The $D\geqslant 6$ MP BHs exist with  arbitrarily large $j$.
Such solutions describe 
\textit{ultra-spinning} BHs,
for which
a wide portion of their horizon is
well approximated by a flat black membrane 
 \cite{Emparan:2003sy}.

The Minkowski spacetime $Q$-balls discussed in Section~\ref{sec_flat}  survive when replacing the near origin region with a MP BH horizon, a feature that has been seen in many contexts within soliton physics.
However, the BH properties are not arbitrary; its angular velocity must equal  the  frequency of the field,
as implied by the synchronization condition (\ref{synch}).                                                                    
 
\begin{figure}[ht]
\hbox to\linewidth{\hss%
	\resizebox{8cm}{6cm}{\includegraphics{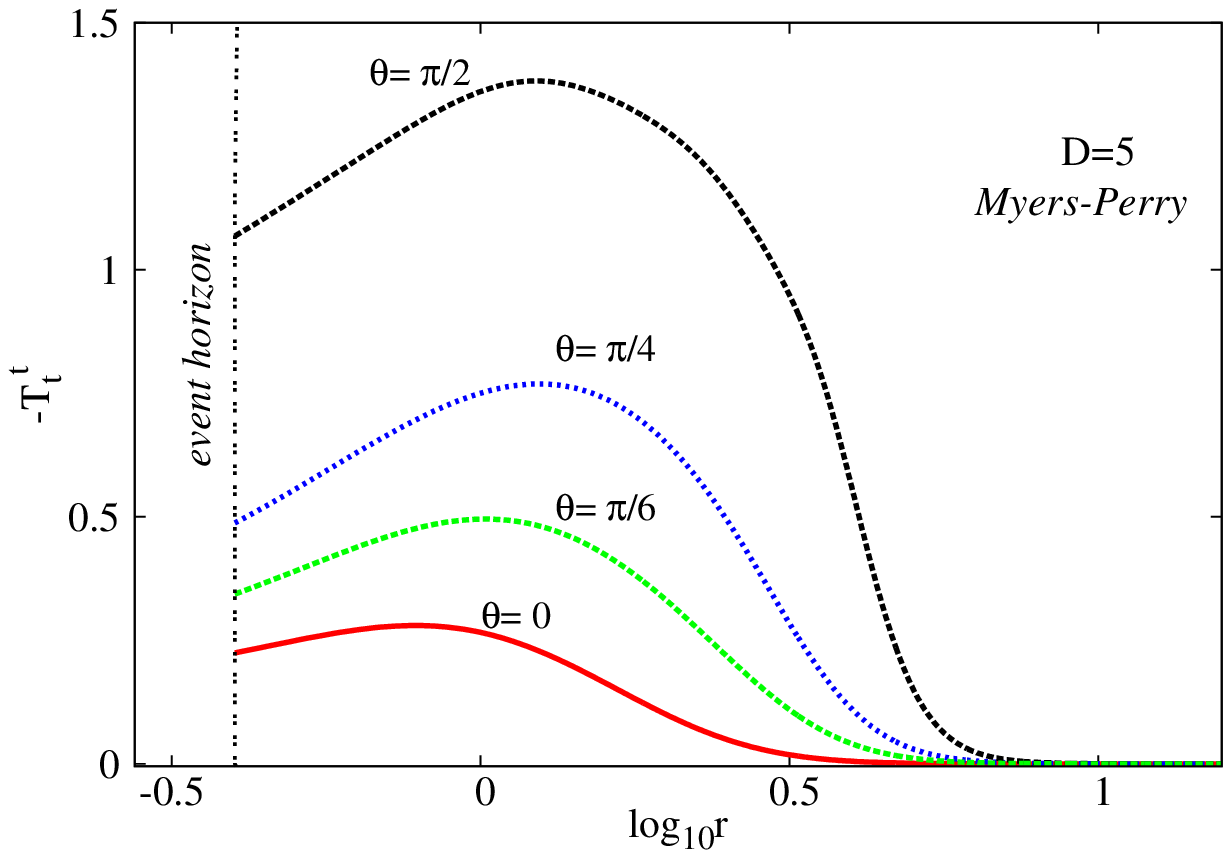}}
\hspace{10mm}%
	\resizebox{8cm}{6cm}{\includegraphics{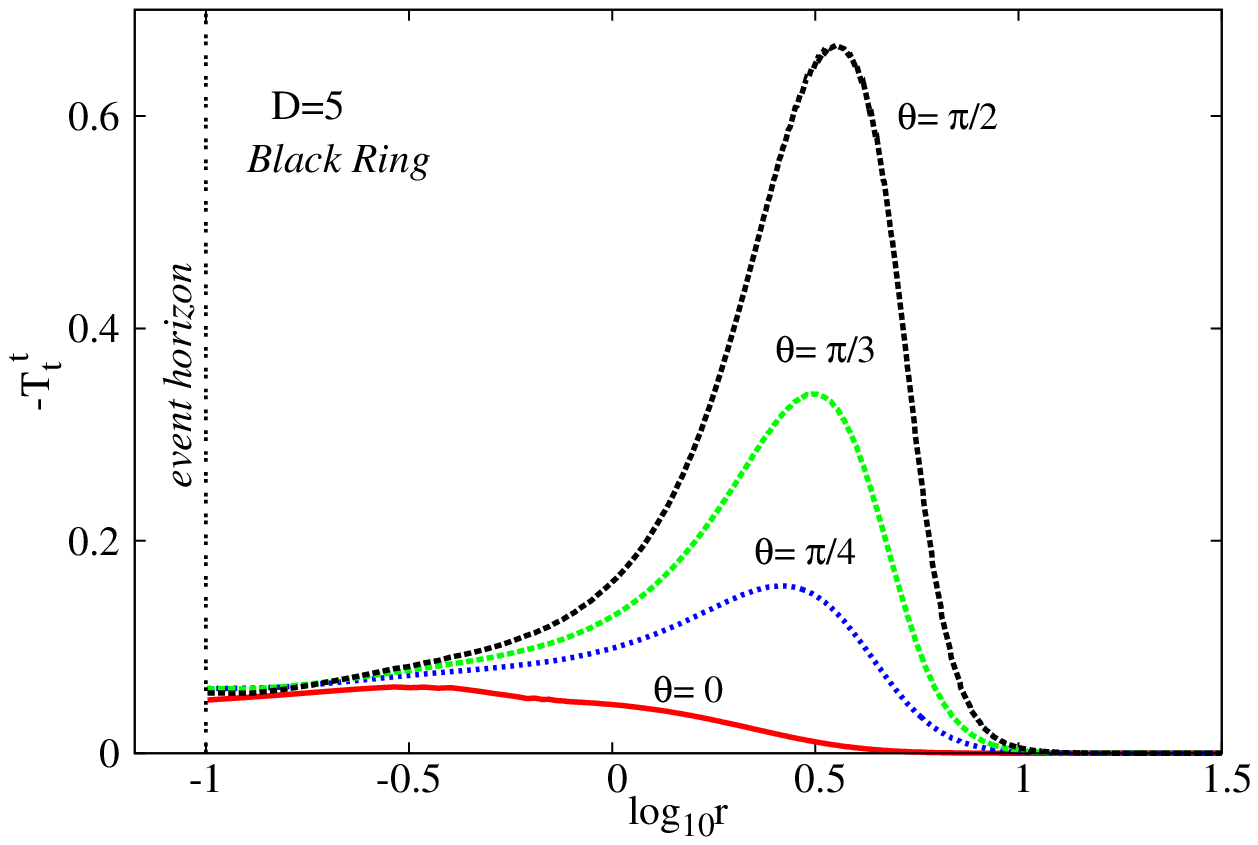}}
\hss}
\hbox to\linewidth{\hss%
		\resizebox{8cm}{6cm}{\includegraphics{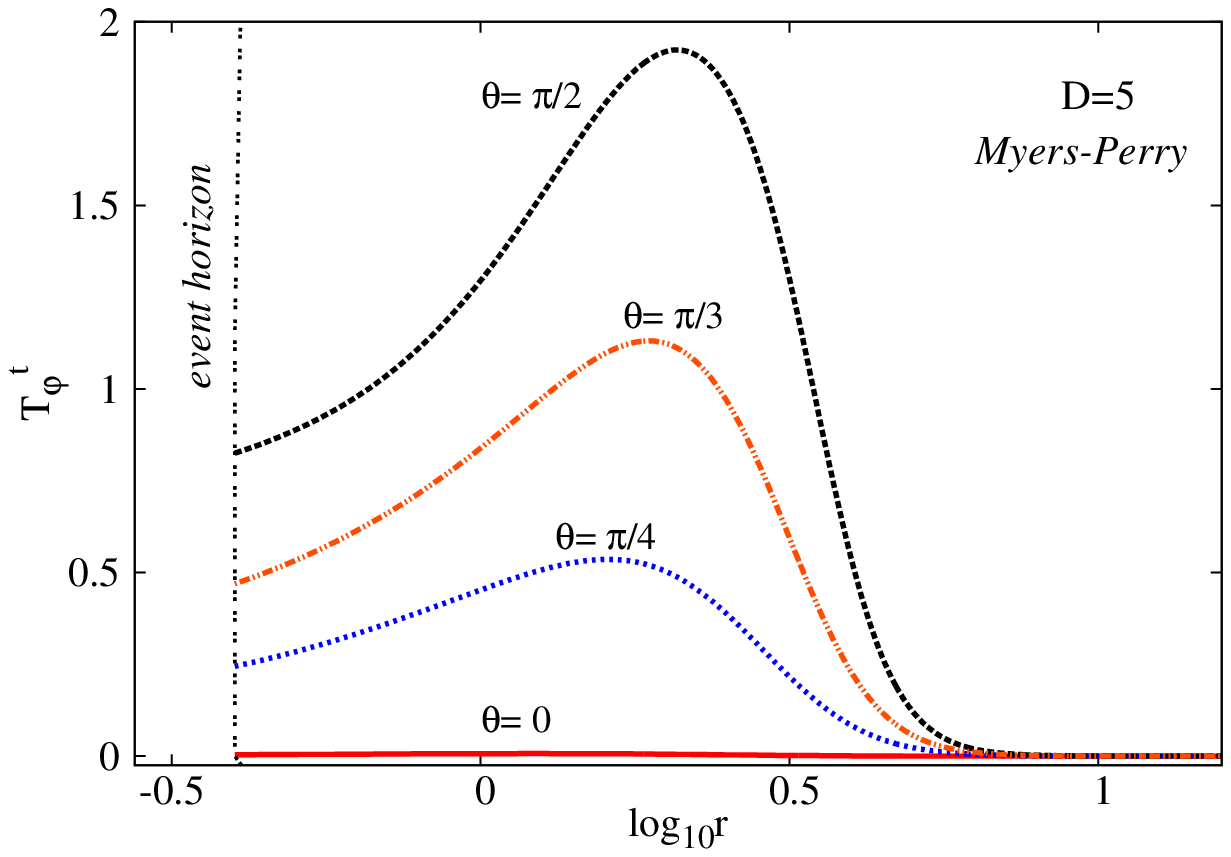}}
\hspace{10mm}%
	\resizebox{8cm}{6cm}{\includegraphics{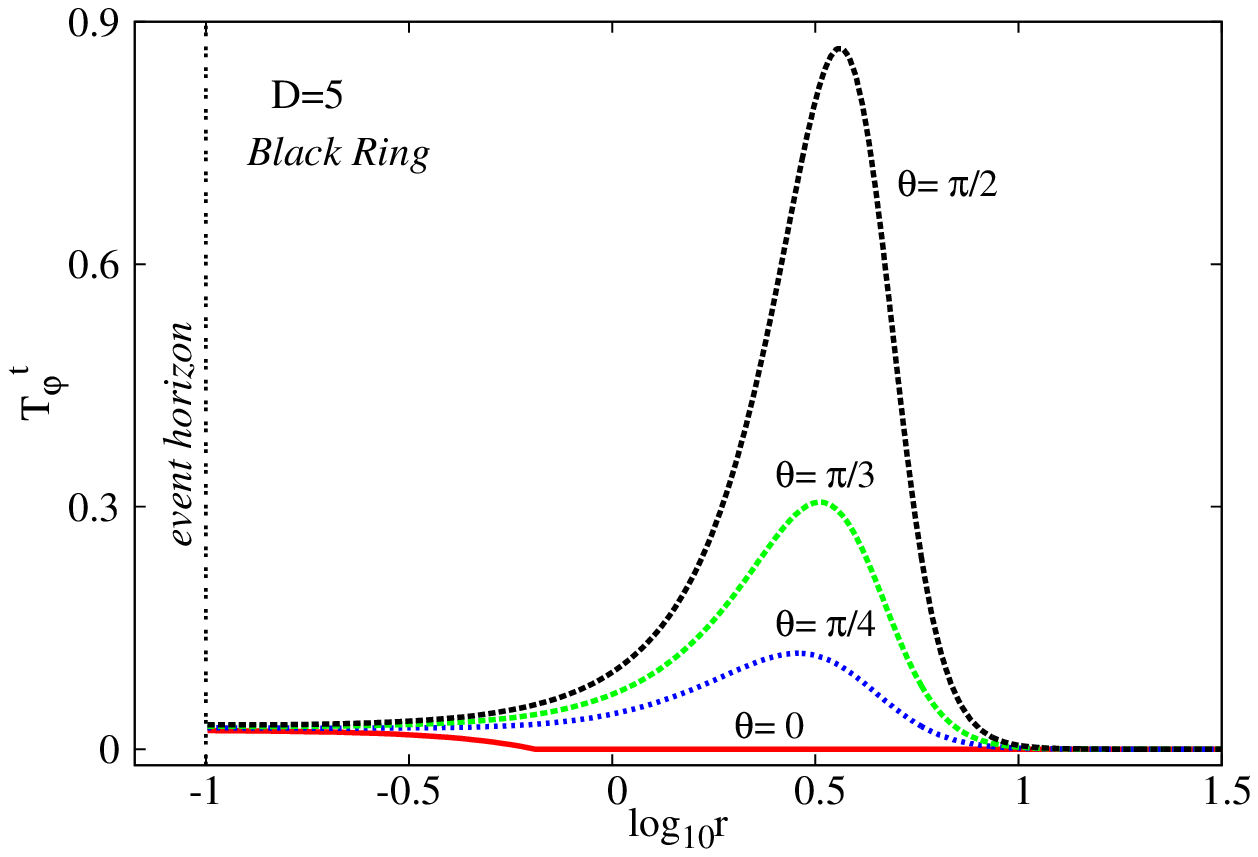}}
\hss}
\caption{\small  
The energy  and angular momentum
density profiles are shown for the same $Q$-clouds as in Figure \ref{profile-MP1}.
}
\label{profile-MP2}
 \end{figure}

For computing such $Q$-balls in the MP background it is convenient to 
introduce a new radial coordinate $R=\sqrt{r^2-r_H^2}$ such that the horizon is now located at $R=0$. 
Using this new radial coordinate, the Klein-Gordon equation~(\ref{KG}) is solved with the following boundary conditions $\partial_R\phi|_{R=0}=\phi|_{\theta=0 }=\partial_\theta \phi|_{\theta=\pi/2}=\phi|_{r=\infty}=0$. 
To obtain solutions, the input parameters are not only $(w,m)$, like in the Minkowski case, but also the event horizon radius, $r_H$. The second free parameter of the MP solution, $a$, is fixed by the choice of $(w,m,r_H)$ and the synchronisation condition\footnote{This implies that for a given $w$, there is a maximal value of $r_H$, with
$r_H^{(max)}=
{m}/{(2w)}$.
 }
(\ref{synch}):  $
{a m}/(a^2+r_H^2)=w$.

Following this strategy, we have performed thorough scans of the domain of existence of $Q$-balls on the $D=5,6$ MP backgrounds.
Partial results for the $D=7$ case were also obtained suggesting a qualitatively
similar picture to $D=6$. These results refer to the choice $m=1$, the only case we have studied in a systematic way, even though some solutions for higher values of $m$ were also obtained. 
The spatial profile of a typical solution is shown in Figure \ref{profile-MP1}, whereas its energy and angular momentum densities are shown in Figure \ref{profile-MP2} (left panels).

To describe the higher dimensional $Q$-clouds' properties, let us recall the $D=4$ case -- $Q$-clouds on the Kerr solution~\cite{Herdeiro:2014pka}. In that case, adding a small BH at the center of the spinning $D=4$ Minkowski space $Q$-balls did not perturb too much
the scalar field distribution, provided the scalar field and horizon co-rotate according to (\ref{synch}). Then,  solutions exist for BHs with
a range of horizon velocities $\Omega^{min}_H<\Omega_H<\Omega^{max}_H(M)$, $i.e.$ in  analogy with the $Q$-balls case. However, the quantitative value of the higher frequency limit depends on the mass of the Kerr BH. At the higher frequency end, the $Q$-balls become diluted, low amplitude and stop to exist at the \textit{existence line} wherein stationary scalar clouds of a massive but non-self-interacting scalar field exist. This is a particular feature of $D=4$ which one does not expect for $D\geqslant 5$ MP BHs, since the latter do not possess superradiant unstable modes of a scalar field. On the low frequency end, solutions cease to exist at a mass independent minimal frequency; as $w\to w_{min}$ both the energy and angular momentum of the $Q$-clouds take very large values (they appear to diverge) making the study of this limit challenging.

\bigskip 

In constructing the $Q$-clouds in $D\geqslant 5$, the choice of the scalar field input parameters, $w,m$, fix $\Omega_H$ via~(\ref{synch}). Then one needs a  second parameter 
to characterize the background geometry (\ref{metric-g}). 
Following \cite{Herdeiro:2014pka}, we take it to be the BH mass $M$.
Then $Q$-cloud solutions may exist for some range of $M$
starting with $M=0$ (flat space) and up to $M=M^{(max)}$.
For $D=4,5$, 
the maximal value of $M$ corresponds to
the set of extremal BHs,
with $M^{(max)}_{D=4}=1/(2\Omega_H)$
and  
$M^{(max)}_{D=5}=3\pi/(8\Omega_H^2)$, but these extremal solutions are regular
for $D=4$ only.
For $D\geqslant 6$,
the solutions with a fixed $\Omega_H$ have a maximal mass 
given by
$
 M^{(max)}_{D\geqslant 6}=p(D)/{\Omega_H^{D-3}}
$
where 
$p(D)= 
  (D-2)(D-3)
\left[
(D-3)(D-5)
\right]^{(D-5)/2}
V_{(D-2)}
/
[
{2^D \pi (D-4)^{D-4}}
]$. 
 For $D=4$, and for $\Omega_H>0.953$, however, not all range of $M$ is realized, as solutions hit the existence line, where they trivialise.  
For $D\geqslant 5$ $Q$-clouds, on the other hand, the solutions do not trivialize for a particular set of background geometries
and, for any given $\Omega_H$, the $Q$-clouds exist for the full allowed range of $M$.
At the ($M$ independent) maximal frequency $w\to w_{max} $, a particular set of $Q$-clouds with finite, nonzero values of the energy $E_{(\Phi)}$ and angular momentum $J_{(\Phi)}$ is reached:  the BH horizon regularizes the flat spacetime divergences of  $Q$-balls at  $w_{max} $.

\begin{figure}[t!]
\hbox to\linewidth{\hss%
	\resizebox{8cm}{6cm}{\includegraphics{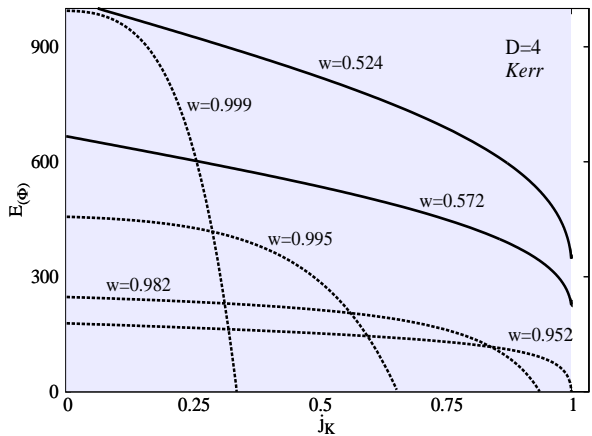}}
\hss}
\caption{\small 
$Q$-clouds energy, $E_{(\Phi)}$,  for several frequencies
as a function of the reduced angular momentum $j_K$, of the $D=4$ Kerr background.
Here and in Figures 
\ref{Ej-MP5},
\ref{Ej-BR} 
the shaded area corresponds to the domain of existence of the 
 $Q$-clouds. 
}
\label{Ej-Kerr}
\end{figure}

\begin{figure}[t!]
\hbox to\linewidth{\hss%
	\resizebox{8cm}{6cm}{\includegraphics{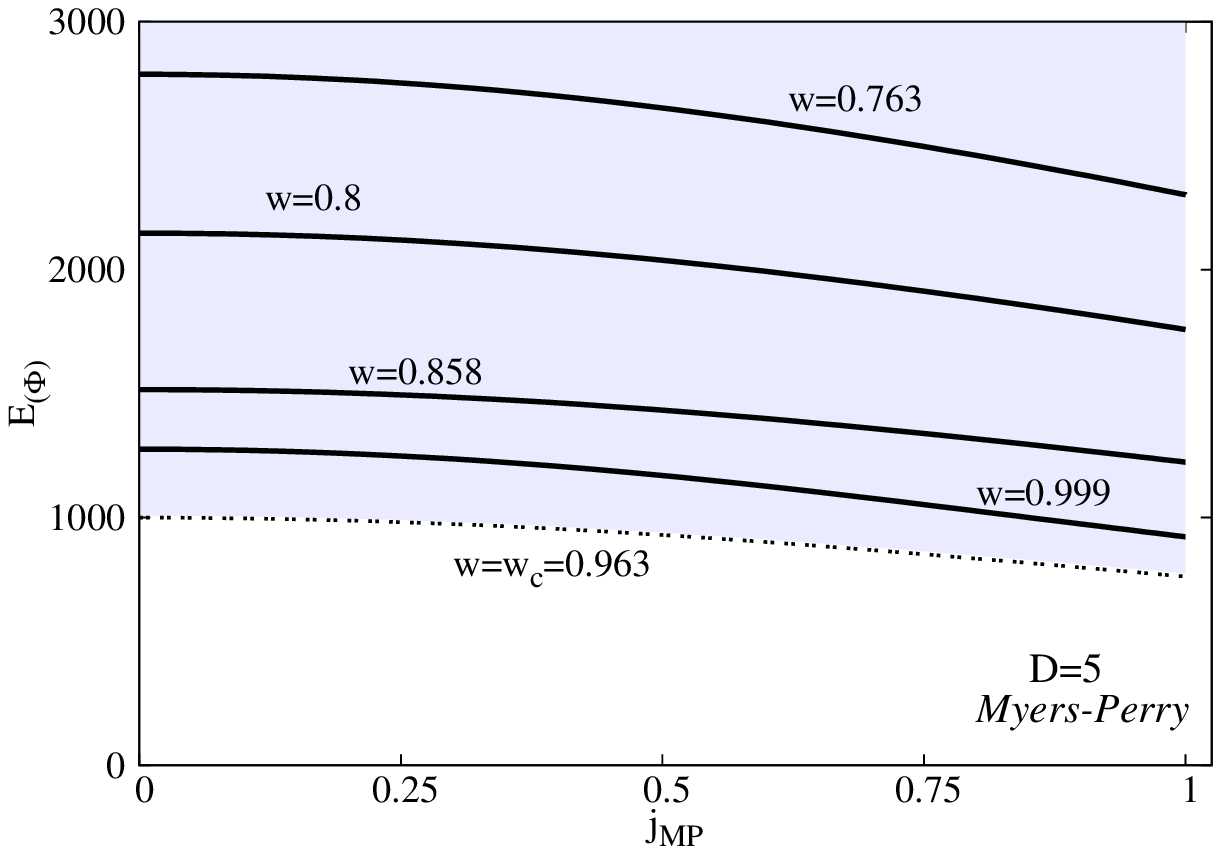}}
	\resizebox{8cm}{6cm}{\includegraphics{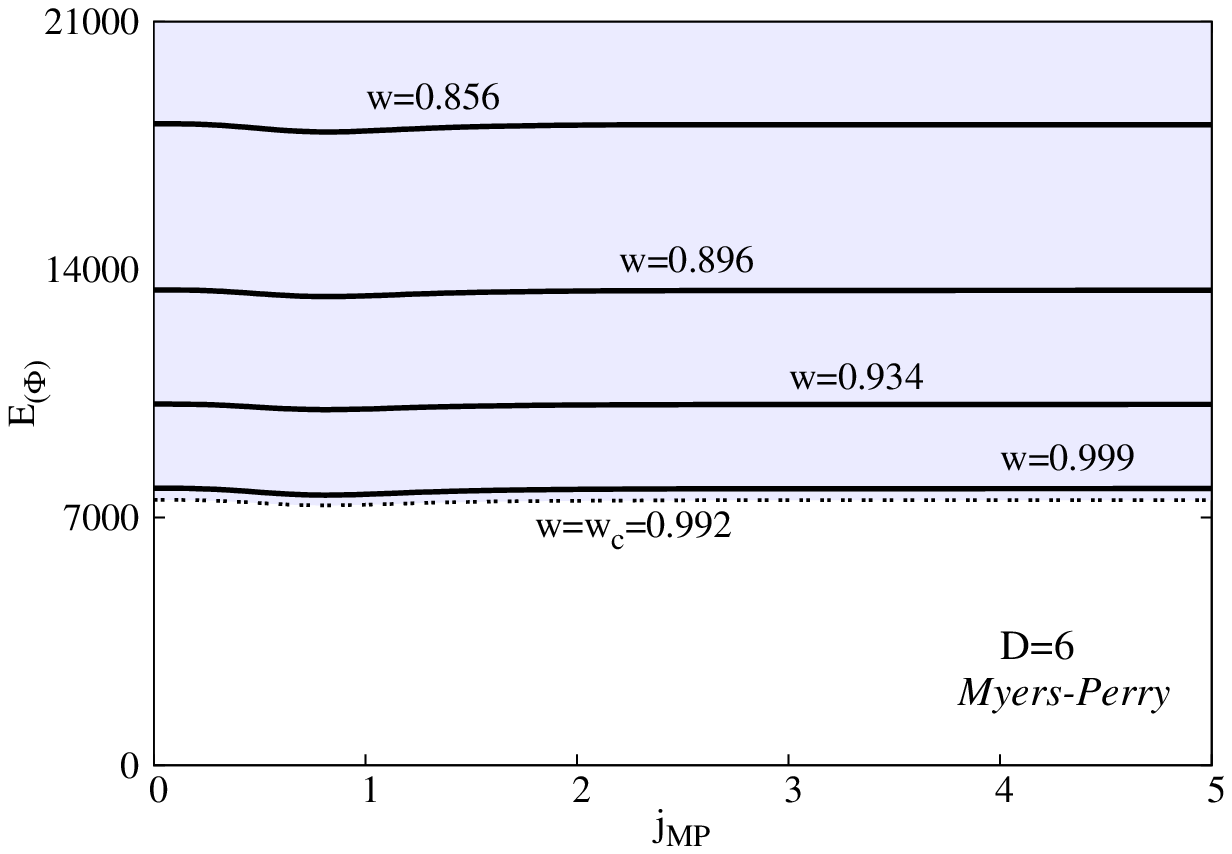}}
\hss}
\caption{\small 
$Q$-clouds energy, $E_{(\Phi)}$,  for several frequencies
 as a function of the reduced angular momentum $j_{MP}$, of the $D=5$ (left panel) and $D=6$ (right panel)  MP backgrounds.  
}
\label{Ej-MP5}
\end{figure}

Despite this distinction between the $D=4$ and $D\geqslant 5$ cases, we remark that  
for any $D$,
the $Q$-clouds exist for the full range of the 
reduced angular momentum of the Kerr/MP backgrounds,
see Figures  \ref{Ej-Kerr}-\ref{Ej-MP5}. 
For $D=5,6$ one observes the existence of a mass/angular momentum gap,
the minimal values for $E_{(\Phi)}$ and $J_{(\Phi)}$
being found for the corresponding critical frequency of the flat spacetime
$Q$-balls.

 {\small \hspace*{3.cm}{\it  } }

 {\small \hspace*{3.cm}{\it  } }

\subsection{$Q$-clouds on a black ring background}

The next application pertains the construction of $Q$-balls on a black ring (BR) background.
We shall focus on the $D=5$ vacuum BR that was found as an exact (closed form solution) in~\cite{Emparan:2001wn}
and represents an asymptotically flat configuration which  possesses a 
horizon topology $\mathcal{S}^2 \times \mathcal{S}^1$. It can be written in the form (\ref{metric-g}), with $\Delta(r)=r^2$ and the corresponding metric functions $F_i,W$ are given in Appendix B
of~\cite{Kleihaus:2014pha}. $D>5$ BRs also exist, but are only known as numerical solutions~\cite{Kleihaus:2012xh,Dias:2014cia}.

In our approach, which can be applied for any $D$, 
the BR solution possesses two input constants:
the event horizon radius $r_H>0$, 
and $R\geqslant r_H$, which provides a rough measure of the ring's $\mathcal{S}^1$ on the horizon.
The $\mathcal{S}^1\times \mathcal{S}^{D-2}$ horizon  topology follows from the behaviour of the metric functions
$F_2$,
$F_3$
at $\theta=0$.
That is the conditions 
$F_2= \partial_\theta F_3=0$
are satisfied for $r_H<r\leqslant R$,
while
$\partial_\theta F_2= F_3=0$, 
for $r>R$ \cite{Kleihaus:2014pha}.

 {\small \hspace*{3.cm}{\it  } }
\begin{figure}[t!]
\hbox to\linewidth{\hss%
	\resizebox{8cm}{6cm}{\includegraphics{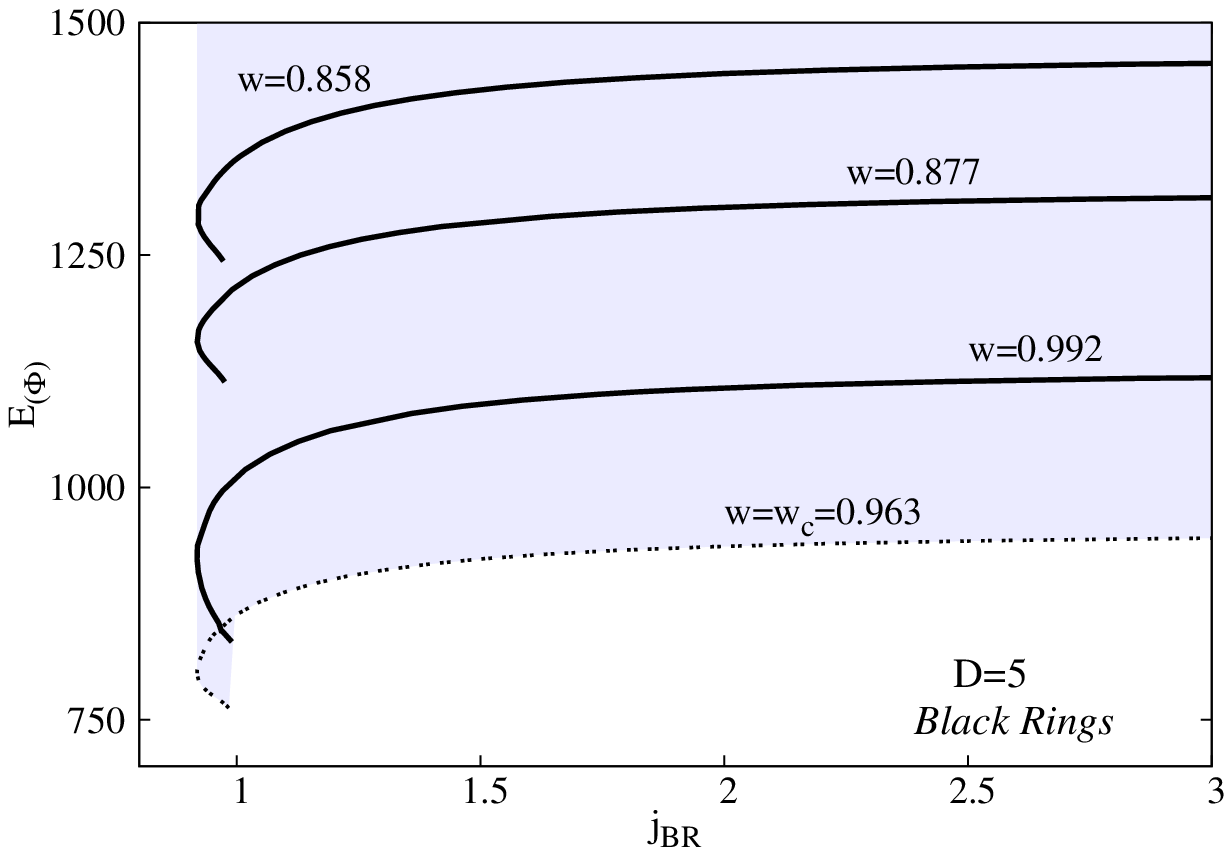}}
		\resizebox{8cm}{6cm}{\includegraphics{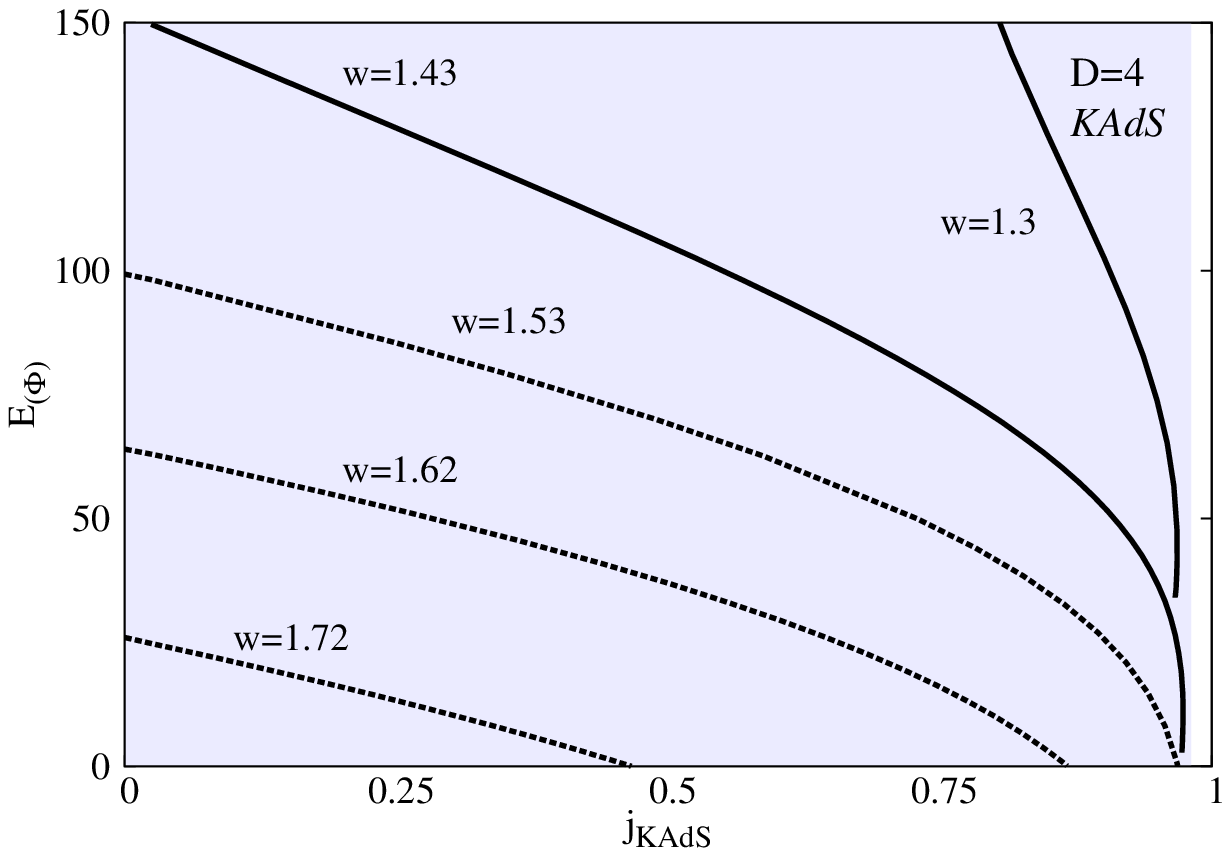}}
\hss}
\caption{\small 
$Q$-clouds energy, $E_{(\Phi)}$ for several frequencies
as a function of the reduced angular momentum  of the $D=5$ black ring backgrounds (left panel) and 
 $D=4$ Kerr-$AdS$ backgrounds  (right panel).  
}
\label{Ej-BR}
\end{figure}

The $Q$-clouds on a BR background are found again, by solving a boundary value problem,
 with the following boundary conditions  at the horizon, infinity and $\theta=\pi/2$: 
$
\partial_r\phi|_{r=r_H}=
\phi|_{r=\infty}=0,  
\partial_\theta \phi|_{\theta=\pi/2}=0$.
The boundary condition  at $\theta=0$
are more complicated,
reflecting the fact that the topology of a $r=const.$
surface changes for some critical value of $r$
\cite{Kleihaus:2014pha}.
 One imposes $
 \partial_\theta \phi|_{\theta=0}=0$, for $r_H<r \leqslant R$, and  $\phi|_{\theta=0}=0$ for $r> R$.
Similarly to the MP case,
these conditions follow  from the construction 
of an approximate form of the solution on the 
boundaries of the domain of integration, together with regularity requirements.

In our approach, 
apart from the winding number $m$
(with $m=1$ for all solutions studied so far) 
we fix the value  of the
  radii $r_H$ and $R$.
	Then, the value of $w$ follows from the synchronization condition (\ref{synch}),
	with \cite{Kleihaus:2014pha}
$
\Omega_H={R(R^2-r_H^2)}[\sqrt{2}(R^2+r_H^2)\sqrt{R^4+r_H^4}]^{-1}\, .
$
In this way, we have scanned the full parameter space of
solutions. 
A typical $Q$-cloud solution is shown in Figures~\ref{profile-MP1}, \ref{profile-MP2} (right panels).

In Figure \ref{Ej-BR}
we show the total energy of the
$Q$-clouds as a function of the reduced angular momentum 
of the background,
$j_{BR}$,  
for several values of the frequency $w$.
The basic features of the solutions can be summarized as follows.
Firstly, the $Q$-clouds exist for 
a set of field frequencies $w_{\rm min}<w<\mu$,
where $w_{\rm min}$ appears to coincide
with the corresponding flat spacetime value.
Both $E_{(\Phi)}$ and $J_{(\Phi)}$ 
stay finite as $w\to \mu$,
while they take very large values (likely, they diverge) as the minimal frequency is approached.
Secondly, one notices the presence of a mass gap:
the $Q$-clouds never trivialize
and we do not find any indication for the occurrence of an  
{\it existence line}.
This is consistent with the absence of a superradiant instability for the $D=5$ vacuum BR.\footnote{
The situation could be different for BRs with two spins (and thus a rotating 2-sphere),
which may inherit the superradiant instability of the Kerr BH~\cite{Dias:2006zv}. 
}
Thirdly, $Q$-clouds exist for 
all allowed values  of the BR's reduced angular momentum,
in particular for the branch of rings connected to MP BHs.
Fourthly, the large-$j_{BR}$ limit is rather subtle.
For a fixed frequency ($i.e.$ angular  velocity of the horizon), this limit is approached as $r_H\to 0$
($i.e.$ vanishing horizon size),
while the radius of the ring $R$ is finite, with $R=m/(\sqrt{2}w)$.
This results in a nonstandard parametrization of flat spacetime  
\cite{Kleihaus:2014pha},
and the  set of
$Q$-balls 
on Minkowski spacetime is recovered in this limit.

\subsection{$Q$-balls on a Kerr-$AdS_4$ background}

Our final example concerns the $D=4$ Kerr-$AdS$ BH.
Although this solution can also 
be written in the generic form (\ref{metric-g}),
it is usually written in the following form
\begin{eqnarray}
ds^{2}  =  -\frac{\Delta_{r}}{\rho^{2}}\left(dt-\frac{a\sin^{2}\theta}{\Xi}d\varphi\right)^{2}+\frac{\Delta_{\theta}\sin^{2}\theta}{\rho^{2}}\left(adt-\frac{r^{2}+a^{2}}{\Xi}d\varphi\right)^{2}
 +\rho^{2}\left(\frac{dr^{2}}{\Delta_{r}}+\frac{d\theta^{2}}{\Delta_{\theta}}\right) \ ,
\end{eqnarray}
where $\Delta_{r}\equiv \left(r^{2}+a^{2}\right)\left(1-{\Lambda r^{2}}/{3}\right)-2\mathcal{M} r$, $\Delta_{\theta}\equiv 1+{\Lambda a^{2}\cos^{2}\theta}/{3}$, $\rho^{2}\equiv r^{2}+a^{2}\cos^{2}\theta$, $\Xi\equiv 1+{\Lambda a^{2}}/{3}$,  $(\mathcal{M} ,a)$ are two input parameters, 
with $\Lambda=-{3}/{L^2}$,
while
  $L$ is the $AdS$ length scale.
The horizons are given by the roots of $\Delta_{r}(r)=0$, $r_H$ being the
largest root of this equation (note that this fixes $\mathcal{M}$ as a function of $(r_H,a)$).
Written in this form, the spacetime is rotating at infinity,
with an angular velocity \cite{Caldarelli:1999xj}
%
$
\Omega_\infty =-\frac{a}{L^2}~.
$
The
relevant quantities of the Kerr-$AdS$ BH are $M={\mathcal{M} }/{\Xi^{2}}$, $J=aM=a{\mathcal{M} }/{\Xi^{2}}$, 
$\Omega_H={a \Xi}/(a^2+r_H^2)$.
We remark that $j=J/M^2<1$ for $AdS$ BHs.

The basic properties of the corresponding $AdS$ $Q$-balls are discussed in a more general context in
\cite{Radu:2012yx}.
The solutions exist, again, for $w_{\rm min}<w<w_{\rm max}$, where
this time
\begin{eqnarray}
\label{wmax}
w_{\rm max}=\frac{m+\Delta_+}{L},~~{\rm with}~~\Delta_+=\frac{3}{2}\left(1+\sqrt{1+\frac{4}{9}\mu^2L^2} \right).
\end{eqnarray}
As $w\to w_{\rm max}$, the mass and angular momentum of the $Q$-balls vanish, while, at least
for the considered parameters  in the potential, both $E_{(\Phi)}$ and  $J_{(\Phi)}$ still seem to diverge as $w\to w_{\rm min}$.
Consequently, $AdS$ works as a regulator at the higher frequency end (just like a BH horizon) and the mass/angular momentum gap found for a Minkowski spacetime background is absent for
$\Lambda<0$. 

Following the general pattern we have been describing, it is possible to add a Kerr-$AdS$ BH in the
center of these solitons.
The construction of the corresponding $Q$-clouds is straightforward,
being similar to that described in~\cite{Herdeiro:2014pka}
for $\Lambda=0$. In particular, one uses the same set of boundary conditions.
As expected, the scalar field synchronizes with the horizon
and not with the angular velocity at infinity $\Omega_\infty$.

Numerical results are shown in Figure \ref{Ej-BR} (right panel) 
for $\Lambda=-0.25$. 
The basic features of the asymptotically flat solutions are recovered.
$Q$-clouds exist in an interval $w_{\rm min}< w<w_{\rm max}$. In the range $w_c\leqslant w<w_{\rm max}$,
$Q$-balls vanish when 
approaching a critical set of Kerr-$AdS$ BHs.
These geometries form the corresponding {\it existence line}. 
$Q$-balls with  $w_{\rm min}<w<w_{c}$ 
exist instead for the full set of Kerr-$AdS$ BHs with a horizon angular velocity given by (\ref{synch}).
In particular,  $Q$-clouds should exist as well for extremal 
 Kerr-$AdS$ BHs, even though we have so far only considered near extremal geometries.
The profile of a typical solution looks rather similar to that found 
in the asymptotically flat case and therefore we shall not display it here.

\section{Final remarks}
\label{sec5}
The existence of $Q$-clouds on a  bald  BH background, as finite energy, regular solutions on and outside the event horizon, establishes that these BHs can support synchronised scalar hair of a self-interacting scalar field, at least in the neighbourhood of the bald BH. Moreover, when considering a self-gravitating scalar field, hence back reacting on the geometry, it has always been the case that the scalar field self-interactions are not mandatory any longer; the non-linearity of gravity yields similar effects even for a linear scalar field. The standard example are $Q$-balls. For flat spacetime solitonic solutions, self-interactions are mandatory. 
But when they self-gravitate in Einstein's gravity (becoming \textit{boson stars}~\cite{Schunck:2003kk}) 
the self-interactions are no longer mandatory and qualitatively similar solutions arise even for a massive scalar field with no self-interactions. Consequently, the existence of $Q$-clouds on a BH background establishes not only the bifurcation towards a family of hairy BHs with self-interacting synchronised hair, but also \textit{suggests} the existence of a family of hairy BHs with massive, but non-self interacting, synchronised hair. 

A different question is if the bifurcation is global or local. In the case of $D\geqslant 5$ MP BHs rotating in a single plain, the nonexistence of a boundary of the $Q$-clouds domain of existence corresponding to an ``existence line" is evidence that the bifurcation is only local. The same applies to the $D=5$ black ring. By contrast, in the Kerr-$AdS$ case, the occurrence of an existence line as one of the boundaries of the $Q$-clouds domain of existence is evidence that the bifurcation is global. 

The results herein further support the generality of the synchronisation mechanism to endow~\textit{any} rotating BH 
with bosonic synchronised hair. Going beyond the paradigmatic example in~\cite{Herdeiro:2014goa} ($D=4$, 
asymptotically flat, with minimally coupled non-self-interacting scalar hair, in Einstein's gravity, without electric charge and with spherical horizon topology), the mechanism also works: ${\bf (i)}$ in different spacetime dimensions $D$, $e.g.$ $D\geqslant 5$~\cite{Brihaye:2014nba,Herdeiro:2015kha}, and $D=3$~\cite{Ferreira:2017cta}; ${\bf (ii)}$ for different asymptotics, $e.g.$ asymptotically $AdS$~\cite{Dias:2011at};  ${\bf (iii)}$ for different matter content, $e.g.$ self-interacting scalar fields~\cite{Kleihaus:2015iea,Herdeiro:2015tia} or vector fields~\cite{Herdeiro:2016tmi,Herdeiro:2017phl}; ${\bf (iv)}$ in modified gravity~\cite{Kleihaus:2015iea}; ${\bf (v)}$ and including electric charge~\cite{Delgado:2016jxq}. Herein we also provide evidence, with the black ring example, that it works for, ${\bf (vi)}$, different horizon topologies. A further test of this generality is the asymptotically \textit{de Sitter} case, for which no such solutions are known and attempts to construct them have, so far, been unsuccessful. $Q$-clouds may be a useful tool to help clarifying this open question.

\vspace{0.5cm}

\section*{Acknowledgements}
J.K. and B.S. would like to acknowledge support by the DFG Research Training Group 1620 "Models of Gravity".
 C.H. and E.R. acknowledge  funding  from  the  FCT-IF  programme,
their work being also
partially supported by the H2020-MSCA-RISE-2015 Grant No.   StronGrHEP-690904, the H2020-MSCA-RISE-2017 Grant No. FunFiCO-777740  and  by  the  CIDMA  project UID/MAT/04106/2013. 
B.S.  acknowledges  funding  from the SAME Dikti grant and ITS local grant 849/PKS/ITS/2017.
The authors  would also  like  to  acknowledge networking support by the COST Action GWverse CA16104.

 \begin{small}

 \end{small}

\end{document}